\documentclass[%
 reprint,
 amsmath,amssymb,
 aps,
]{revtex4-2}

\usepackage{graphicx}
\usepackage{dcolumn}
\usepackage{bm}

\usepackage[english]{babel}
\usepackage[utf8x]{inputenc}
\usepackage[T1]{fontenc}
\usepackage{braket}
\usepackage{mathtools}
\usepackage{flushend}

\usepackage{amsmath}
\usepackage{siunitx}
\usepackage{graphicx}
\usepackage{dcolumn}
\usepackage{bm}
\usepackage{array}
\usepackage[shortlabels]{enumitem}
\newcolumntype{?}{!{\vrule width 1pt}}
\usepackage[colorinlistoftodos]{todonotes}
\usepackage[colorlinks=true,linkcolor=blue,citecolor=blue,urlcolor=blue]{hyperref}
\usepackage{slashed}

\usepackage{bibunits}
\defaultbibliographystyle{apsrev4-1}
\usepackage{etoolbox}

\makeatletter
\newcommand*{\newbibstartnumber}[1]{%
  \apptocmd{\thebibliography}{%
    \global\c@NAT@ctr #1\relax
    \addtocounter{NAT@ctr}{-1}%
  }{}{}%
}
\makeatother

\newcommand\bb[1]{\mbox{\boldmath{$#1$}}}

\newcommand\ROS{\bb{bb}\,\bb{:} \,\bb{\nabla}\bb{u}}

\renewcommand\prl{Phys.~Rev.~Lett.}
\newcommand\rpp{Rep.~Prog.~Phys.}
\newcommand\physrep{Phys.~Rep.}
\newcommand\araa{Ann.~Rev.~Astron.~Astrophys.}
\newcommand\mnras{Mon.~Not.~Roy.~Astron.~Soc.}
\newcommand\jopp{J.~Plasma Phys.}
\newcommand\apjl{Astrophys.~J.~Lett.}
\newcommand\pnas{Proc.~Nat.~Acad.~Sci.}
\newcommand\prr{Phys.~Rev.~Research}
\newcommand\natp{Nat.~Phys.}
\renewcommand\pop{Phys.~Plasmas}
\newcommand\jetp{Soviet J.~Exp.~Theor.~Phys.}
\makeatletter
\let\cat@comma@active\@empty
\makeatother
\begin{document}


\title{Magnetogenesis in a collisionless plasma: from Weibel instability to turbulent dynamo}

\author{Muni Zhou}
\thanks{\href{mailto:munizhou@princeton.edu}{munizhou@princeton.edu} 
	\vspace{0.5cm}}
\affiliation{Department of Astrophysical Sciences$,$ Princeton University$, $ Peyton Hall$, $ Princeton$,$ NJ 08544$,$ USA}
\affiliation{School of Natural Sciences$,$ Institute for Advanced Study$, $ Princeton$,$ NJ 08544$,$ USA}
\author{Vladimir Zhdankin}
\affiliation{Center of Computational Astrophysics$,$ Flatiron Institute$, $ 162 5th Avenue$, $ New York$,$ NY 10010$,$ USA}
\author{Matthew W.~Kunz}
\affiliation{Department of Astrophysical Sciences$,$ Princeton University$, $ Peyton Hall$, $ Princeton$,$ NJ 08544$,$ USA}
\affiliation{Princeton Plasma Physics Laboratory$,$ PO Box 451$,$ NJ 08544$,$ USA}
\author{Nuno F.~Loureiro}
\affiliation{Plasma Science and Fusion Center$,$ Massachusetts Institute of Technology$,$ Cambridge$,$ MA 02139$,$ USA}
\author{Dmitri A.~Uzdensky}
\affiliation{Center for Integrated Plasma Studies$,$ Physics Department$,$ UCB-390$,$ University of Colorado$,$ Boulder$,$ CO 80309$,$ USA}

\date{\today}

\begin{abstract}
We report on a first-principles numerical and theoretical study of plasma dynamo in a fully kinetic framework. By applying an external mechanical force to an initially unmagnetized plasma, we develop a self-consistent treatment of the generation of ``seed'' magnetic fields, the formation of turbulence, and the inductive amplification of fields by the fluctuation dynamo. Driven large-scale motions in an unmagnetized, weakly collisional plasma are subject to strong phase mixing, which leads to the development of thermal pressure anisotropy. This anisotropy triggers the Weibel instability, which produces filamentary ``seed'' magnetic fields on plasma-kinetic scales. The plasma is thereby magnetized, enabling efficient stretching and folding of the fields by the plasma motions and the development of Larmor-scale kinetic instabilities such as the firehose and mirror.
The scattering of particles off the associated microscale magnetic fluctuations provides an effective viscosity, regulating the field morphology and turbulence. During this process, the seed field is further amplified by the fluctuation dynamo until they reach energy equipartition with the turbulent flow. By demonstrating that equipartition magnetic fields can be generated from an initially unmagnetized plasma through large-scale turbulent flows, this work has important implications for the origin and amplification of magnetic fields in the intracluster and intergalactic mediums. 
\end{abstract}

 \maketitle


\section{Introduction}
\label{sec:intro}
The origin and evolution of cosmic magnetic fields is one of the most important long-standing problems in astrophysics and cosmology~\cite{kulsrud2008origin}.
In galaxies and clusters of galaxies, large-scale magnetic fields with up to micro-Gauss strengths are found to be ubiquitous through observations of Faraday rotation, synchrotron emission, and Zeeman splitting~\cite[e.g.,][]{beck1996galactic,carilli2002cluster,bonafede2010coma}.
The amplification of pre-existing ``seed'' magnetic fields by the plasma dynamo---a fundamental plasma process that converts the mechanical energy of plasma motions to magnetic energy through magnetic induction---is believed to be essential in producing such dynamically important magnetic fields. In contexts such as the intracluster medium (ICM) of galaxy clusters, the dynamo is thought to proceed through successive stretching of magnetic fields by (gravitationally driven) chaotic flows, resulting on the average in amplification of the magnetic-field strength.

Plasma dynamos have been studied extensively within a magnetohydrodynamic (MHD) framework~\cite{Subramanian1994,brandenburg2005astrophysical,rincon2019dynamo}, but only recently using a kinetic framework~\cite{rincon2016turbulent,st2018fluctuation,pusztai2020dynamo}, even though a kinetic treatment of the dynamo is important because cosmic plasmas are typically weakly collisional, i.e., the particles’ Coulomb mean free paths are comparable to or even exceed the characteristic macroscopic length scale of an astrophysical system. Under such conditions, the influence of micro-physical plasma processes on the dynamo is potentially significant. For example, hybrid-kinetic (kinetic ions, fluid electrons) studies of the dynamo~\cite{rincon2016turbulent,st2018fluctuation} have shown that micro-scale kinetic instabilities determine the effective viscosity of the turbulent plasma and, in so doing, control the amplification rate of any seed magnetic field. Going further---namely, realizing a self-consistently generated seed field and capturing the influence of collisionless electrons---requires a fully kinetic treatment of the dynamo process. This is the purpose of this paper.

In a weakly collisional plasma, anisotropy in the thermal motions of the particles provides free energy to create magnetic fields from an initially unmagnetized state through the Weibel instability~\cite{weibel1959,zhou2022spontaneous}.
In the post-reionization, unmagnetised universe, turbulent motions at intergalactic scales may trigger plasma instabilities (in particular, the Weibel instability~\cite{weibel1959}), which deplete this thermal free energy to produce a microscopic ``seed'' magnetic field.
As the plasma is magnetized by this seed field, the bulk flow is then able to stretch and fold the magnetic field, increasing its overall strength. 
The approximate conservation of the adiabatic invariant of the magnetic moments $\mu$ by the magnetized particles implies that the growth of the magnetic field will bias the thermal motion of the plasmas with respect to the magnetic-field direction, represented by a thermal-pressure anisotropy~$\Delta$.
This pressure anisotropy serves as a source of free energy for the mirror and firehose plasma instabilities.
The scattering of particles off the Larmor-scale fluctuations driven by these two instabilities plays an important role for the plasma dynamo by controlling the plasma viscosity (and possibly resistivity as well)~\cite{kunz2014firehose,st2018fluctuation}, and by breaking the adiabatic invariance of $\mu$, the conservation of which in the absence of pitch-angle scattering would place a prohibitive constraint on the energy budget for the growth of magnetic fields~\cite{Helander2016}.

In this work, we develop a fully kinetic, self-consistent description of the generation and amplification of magnetic fields in an initially unmagnetized plasma under large-scale chaotic motions.
In Section~\ref{sec:theory}, we discuss our theoretical expectations for four distinguishable phases of the {\em ab-initio} plasma dynamo, as well as the scale separation required between the macroscopic flows and the plasma skin depth to capture these phases in a numerical simulation. 
We then present results from particle-in-cell (PIC) simulations of the plasma dynamo in Section~\ref{sec:numerical}, which we find to be qualitatively consistent with the theoretical expectations.
We conclude in Section~\ref{sec:discission} with a brief discussion on the properties of a fully kinetic dynamo and how our results fit into the broader narrative of cosmic magnetogenesis.

\section{Theoretical expectations}
\label{sec:theory}

In a collisionless plasma, kinetic instabilities often play an essential role in generating seed magnetic fields and regulating the material properties of the plasma, e.g., its effective dynamical viscosity, thermal conductivity, and electrical resistivity \citep{ks19}. The interplay between this microscale physics, and its macroscale consequences for the establishment of turbulent flows and the amplification and sustenance of cosmic magnetic fields, are central ingredients in any predictive theory for the plasma dynamo. 
In this section, we review some of this physics, taking care to distinguish between what has been rigorously established and what is more speculative.

We first define three dimensionless quantities to describe the energetics of the system. 
The first, which quantifies the average magnetic energy density in the plasma, is the (inverse) plasma beta parameter, $\beta^{-1}\equiv \braket{B^2/8\pi}/\braket{P}$, where $P$ is the  plasma pressure and  $\langle\,\cdot\,\rangle$ denotes the domain average. 
The bulk flow energy is quantified using the square of the Mach number, $M^2\equiv (U_{\rm rms}/v_{\rm th})^2$, with $U_{\rm rms}$ being the root-mean-square bulk flow speed and $v_{\rm th} \equiv \sqrt{\braket{P/\rho}}$ being the thermal speed, with $\rho$ the total mass density. 
The free thermal energy of the plasma is represented by the pressure anisotropy, $\Delta \equiv \braket{P_\perp/P_\parallel-1}$, where $P_\perp$ ($P_\parallel$) is the thermal pressure perpendicular (parallel) to the local magnetic field. This definition presumes a magnetized plasma; when the plasma is unmagnetized, the relevant pressure anisotropy is that measured with respect to the axis along which the pressure tensor has its maximum eigenvalue (see~\cite{zhou2022spontaneous} for details).
In the Weibel-seeded plasma dynamo, there are two ways in which pressure anisotropy is produced. 
The first, most relevant to the early unmagnetized stage, issues from the collisionless phase mixing of the driven shear flows~\cite{zhou2022spontaneous}.
Unmagnetized particles carrying the momentum of the local bulk flows have random thermal motions. 
The free streaming of these particles smooths out the spatial variation of the bulk flows and leads to the development of velocity-space anisotropy in particle distributions and thus pressure anisotropy. 
The second means of producing pressure anisotropy requires the plasma to be magnetized, as it relies on the adiabatic invariance of the particles' magnetic moments $\mu\equiv mv_\perp^2/2B$, where $m$ and $v_\perp$ are the mass and perpendicular velocity of each particle, to couple their perpendicular thermal energy to the magnetic-field strength. Consequently, as the bulk flows stretch and amplify the magnetic field, $P_\perp$ increases relative to~$P_\parallel$.
Once produced, $\Delta$ provides a free-energy source for driving rapidly growing kinetic instabilities, predominantly in the form of skin-depth- or Larmor-scale magnetic fluctuations.
The scattering or trapping of particles as they interact with these fluctuations leads to an effective dynamical viscosity (in addition to that caused by phase mixing and particle collisions), which in turn constrains the large-scale flows. 

\begin{figure}[h!]
    \centering
    \includegraphics[width=0.49\textwidth]{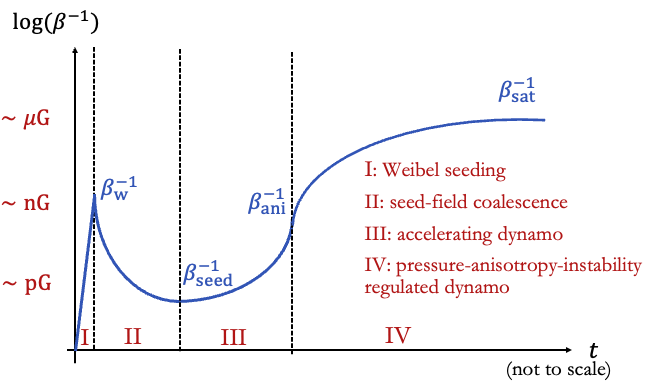}
    \caption{A qualitative illustration of the predicted time evolution of $\beta^{-1}$ (or magnetic energy) in the Weibel-seeded, turbulent plasma dynamo, divided into four main phases. Reference values of the predicted magnetic-field strength given ICM conditions are given on the ordinate. }
    \label{fig:sketch}
\end{figure}

In what follows, we provide estimates for the evolution of $\beta^{-1}$ given driven turbulence characterized by Mach number $M$ and characteristic scale $L$. 
With asymptotically large scale separation between the macroscopic astrophysical flows and the microscopic plasma kinetic scales, we anticipate four main phases of magnetic-field amplification, as illustrated schematically in Figure~\ref{fig:sketch} and detailed in the following subsections.

\subsection{Seeding of magnetic fields by Weibel instability}
\label{sec:theory_weibel}
The Weibel instability has been recognized and widely studied as a mechanism to generate magnetic fields. It is particularly versatile, as its only requirement is pressure anisotropy in an unmagnetized plasma. Despite this versatility, the Weibel instability has been studied mainly in the context of local counter-streaming configurations such as collisionless shocks~\cite[e.g.,][]{medvedev1999generation,spitkovsky2008,Kato2008,medvedev2006cluster} and laser-plasma interactions~\cite[e.g.,][]{schoeffler2014magnetic,huntington2015observation}, and has only recently been considered in the more global context of low-Mach-number turbulence, e.g., in galaxy clusters and the intergalactic medium~\cite{zhou2022spontaneous}. In this paper, we are concerned with the latter case.

\citet{zhou2022spontaneous} presented an analytical and numerical investigation of the development and saturation of the Weibel instability in an electron-positron plasma under the action of a large-scale shear flow (as a local approximation of a turbulent system).
In an initially unmagnetized plasma, any inhomogenous flow is subject to efficient phase mixing via the thermal motions of the particles. 
This phase mixing leads to the development of free energy in the plasma in the form of a pressure anisotropy that increases on a hybrid thermal-dynamic time scale, $\Delta \simeq M (t v_{\rm th}/L)^2$. 
In response to this slowly evolving background, the fast, kinetic-scale electron Weibel instability is triggered, with an instantaneous linear growth rate $\gamma_{\rm w} \simeq \Delta^{3/2}\omega_{{\rm p}e}v_{\rm th}/c$ for the most unstable linear mode $k_{\rm w} \simeq \Delta^{1/2}/d_e$, where $\omega_{{\rm p}e} \equiv \sqrt{4\pi n e^2/m_e}$ is the plasma frequency and $d_e \equiv c/\omega_{{\rm p}e}$ is the electron skin depth.
In consideration of this electron-only (or electron-positron) Weibel instability, $\Delta$ and $M$ in these expressions are defined using the electron thermal speed. 
As $\Delta$ increases slowly, the linear growth rate of the Weibel instability also increases, and the magnetic perturbations grow super-exponentially~\cite{zhou2022spontaneous}.
This rapid growth proceeds until the instability enters a nonlinear regime in which the depletion of the free energy (quantified by~$\Delta$) by the instability balances its replenishment by the persistent phase mixing of the bulk momentum.  
In this nonlinear regime, the spontaneously produced magnetic field continues its growth, but with a decreased growth rate. At some point this field begins to affect the trajectories of the particles and eventually magnetize the plasma. 
The instability then reaches saturation when the electron Larmor radius $\rho_e \equiv m_e v_{\rm th}c/eB$ decreases to become comparable to the coherence scale of the Weibel seed field, $\rho_e k_{\rm w} \sim 1$; this is known as the trapping condition~\cite{davidson1972nonlinear,kato2005saturation}.

The dependence of the saturated magnetic energy (${\propto} \beta^{-1}$) and coherence scale (${\propto} k_{\rm w}^{-1}$) of the Weibel seed field on the scale separation $L/d_e$ and the Mach number were found to be given by~\cite{zhou2022spontaneous}
\begin{equation}
    \beta_{\rm w}^{-1} \sim (L/d_e)^{-1/2} M^{1/4},
    \label{eq:beta_weibel}
\end{equation}
\begin{equation}
    k_{\rm w}d_e \sim (L/d_e)^{-1/4}M^{1/8}.
    \label{eq:k_weibel}
\end{equation}
These results set the expectation that, under the generic motions of astrophysical turbulence, seed magnetic fields are automatically generated through the Weibel instability and plasmas are spontaneously magnetized.
Without making additional assumptions on the origins of seed fields, the $\beta^{-1}_{\rm w}$ given by Eq.~\eqref{eq:beta_weibel} provides a lower bound on the seed magnetic energy for any turbulent dynamo. 

The above discussion only considered the electron Weibel instability. In reality, as the ions develop a pressure anisotropy of their own, they should also become unstable to a Weibel instability, which would produce magnetic fields on the ion-kinetic scales. Because the detailed effects of the magnetized electrons on this ion Weibel instability are still unclear at this time, we simply assume that the magnetic energy and wavenumber of the ion Weibel fields should obey the same scaling as Eqs.~\eqref{eq:beta_weibel}--\eqref{eq:k_weibel}, with $d_e$ replaced by the ion skin depth $d_i$ and $M$ defined using the ion thermal speed. Adopting these replacements would boost the value of $\beta_{\rm w}^{-1}$ at the end of the ion-Weibel stage by a relatively modest factor of $(m_i/m_e)^{3/8} \approx 17$, and the value of $k_{\rm w}$ given by the ion Weibel field would decrease by a factor of $(m_i/m_e)^{5/16} \approx 10$. For the remainder of this section, we assume that there is a phase of ion-Weibel growth of the magnetic field; the numerical experiments presented in Section~\ref{sec:numerical} adopt an electron-position plasma, for which no such phase exists.

\subsection{Inverse cascade of the Weibel seed field}
\label{sec:theory_coalescence}

Because the Weibel field is produced on a time scale much shorter than the flow-crossing time scale ${\sim}L/U_{\rm rms}$, the impact of the large-scale shear flow on its initial evolution is only minimal. 
The Weibel field is thus expected to evolve without directly interacting with the background flow for $t\lesssim L/U_{\rm rms}$.
It has been repeatedly found that these Weibel fields, after their formation and saturation, will coalesce and increase their coherence length~\cite[e.g.,][]{medvedev2004long,Kato2008,zhou2022spontaneous}, that is, inverse-cascade to larger scales.
In order to predict the properties of these fields at the moment they become seeds for a fluctuation dynamo (i.e., at $t\sim L/U_{\rm rms}$), it is essential to know what scale the Weibel fields can reach through coalescence within one flow-crossing time, and how their magnetic energy evolves during this inverse cascade. 
\citet{zhou2019magnetic,zhou2020multi,zhou2021statistical} derived a simplified analytical model based on fundamental conservation laws to describe the evolution of initially small-scale magnetic fields during their successive coalescence~\footnote{Their study was conducted using the resistive-MHD equations, but the conservation laws they used to derive the scalings for the evolution of the system are expected to hold more generally (beyond a fluid description)}.
Those authors identified magnetic reconnection as the key mechanism enabling the growth of magnetic fields' characteristic length scale and setting the associated time scale.
They found that the decay of magnetic energy and the growth of the coherence length of the magnetic fields (or, equivalently, the decrease of its corresponding wavenumber $k$) are described by scalings 
\begin{equation}
    \beta^{-1} \sim \beta_{\rm w}^{-1} (t/\tau_{\rm rec})^{-1}, \  \text{and} \ k \sim k_{\rm w} (t/\tau_{\rm rec})^{-1/2},
    \label{eq:inverse_law}
\end{equation}
respectively. Here $\tau_{\rm rec}\equiv \epsilon_{\rm rec}^{-1} \sqrt{\beta_{\rm w}}/(k_{\rm w} v_{\rm th})$ is the reconnection time scale for the initial fields (i.e., the saturated Weibel fields with energy $\beta_{\rm w}^{-1}$ and wavenumber $k_{\rm w}$) and $\epsilon_{\rm rec}$ is the dimensionless reconnection rate; values of $\epsilon_{\rm rec} \sim 0.1$ are usually found in numerical studies of reconnection in a collisionless, well-magnetized plasma~\cite[e.g.,][]{daughton2007collisionless,daughton2011} (although the detailed physics of magnetic reconnection in the high-$\beta$ regime is still unclear).

Combining Eqs.~\eqref{eq:beta_weibel}--\eqref{eq:inverse_law}, we obtain the energy and inverse length scale of the coalescing Weibel filaments at $t\sim L/U_{\rm rms}$:
\begin{equation}
    \beta^{-1}_{\rm seed} \sim (L/d_i)^{-1}M\epsilon_{\rm rec}^{-1},
    \label{eq:beta_t1}
\end{equation}
\begin{equation}
    k_{\rm seed} d_i \sim (L/d_i)^{-1/2}M^{1/2}\epsilon_{\rm rec}^{-1/2}.
    \label{eq:k_t1}
\end{equation}
In a typical ICM, $M \sim 0.1$ (and so $M \epsilon_{\rm rec}^{-1} \sim 1$) and $L/d_i \sim 10^{14}$.
If the above scalings are correct, the remnant Weibel seed fields at the end of the inverse cascade process would have an energy of only $\beta_{\rm seed}^{-1} \sim 10^{-14}$ and reside on length scales much larger than the kinetic scales, $k_{\rm seed} d_i \sim 10^{-7}$. 
Note that, during this process, the ratio of the particles' Larmor radii ($\rho_i \propto \beta^{1/2} $) and the coherence scale of the magnetic field, {\it viz.},~$\rho_i k \sim \beta^{1/2}_{\rm w} k_{\rm w}$, remains a constant.
That is, after the particles become magnetized at the saturation of Weibel fields, they remain magnetized during the inverse cascade of these fields.

\subsection{Accelerating dynamo}
\label{sec:theory_explosive}
At $t \sim L/U_{\rm rms}$, the large-scale plasma flow is established and the coalescence of the Weibel fields is replaced by the stretching and folding of the field lines by the flow.
The adiabatic conservation of the magnetic moment $\mu$ during this stretching implies the generation of pressure anisotropy. As the pressure anisotropy grows, the magnetized plasma becomes unstable to kinetic plasma instabilities, namely the mirror~\cite{shapiro1964quasilinear,barnes1966collisionless,southwood1993mirror,hellinger2007comment} and the firehose~\cite{rosenbluth1956stability,chandrasekhar1958stability,parker1958dynamical,vedenov1958some,yoon1993effect,hellinger2000new} (the ion cyclotron instability may also play a role when $\Delta > 0$~\cite{riquelme2015}).
The mirror instability, with the threshold $\Delta\gtrsim 1/\beta$, should occur in regions where the field lines are stretched and the field strength increases; the firehose instability, with the threshold $\Delta\lesssim -2/\beta$, should occur in regions where the field lines are bent (which, in the fluctuation dynamo, are statistically also where the field strength has decreased~\cite{schekochihin2004simulations}).

After these instabilities grow and saturate, particles scatter off of the associated Larmor-scale distortions in the magnetic field and isotropize the velocity distribution.
This scattering can be interpreted as an effective collisionality~\cite{kunz2014firehose,riquelme2015}, which we denote as $\nu_{\rm eff}$, and which supplants the (typically slower) Coulomb collisionality~$\nu_{\rm c}$. Following the Chew--Goldberger--Low equations~\cite{chew1956boltzmann}, assuming incompressibility, and taking into account the isotropizing effect of the effective collisions~\cite{schekochihin2006turbulence,Rosin2011}, the evolution of $\Delta$ with $|\Delta| \lesssim 1/\beta \ll 1$ can be  written heuristically as
\begin{equation}
    \frac{{\rm d}\Delta}{{\rm d}t} \approx 3 \frac{ {\rm d}\ln B}{{\rm d}t}-\nu_{\rm eff} \Delta.
    \label{eq:delta_brag}
\end{equation}
This equation states that pressure anisotropy is produced through adiabatic invariance and relaxed by an effective collisionality. 

In a turbulent environment, the spatio-temporal inhomogeneity of the fluctuations, the pressure anisotropy, and thus the effective collisionality complicate a detailed description of the plasma dynamo. What follows in the remainder of this subsection and the next one (\S\ref{sec:theory_dynamo}) is a scenario for the inductive phase of the plasma dynamo, one that is based on a combination of theoretical arguments and results from existing hybrid-kinetic simulations of this phase.

We begin by associating an effective parallel Reynolds number $\text{Re}_\parallel \equiv U_{\rm rms}L/(v_{\rm th}^2/\nu_{\rm eff})$ with the effective collisionality, which for $\nu_{\rm eff}>\nu_{\rm c}$ is larger than the Reynolds number associated with Coulomb collisions.
Because ${\rm Re}_\parallel$ determines the maximum value of the field-parallel rate of strain of the flow, $\ROS$, it also controls the amplification rate of the magnetic field.
Assuming the Kolmogorov scaling~\cite{Kolmogorov1941} for the turbulent flow (which is not {\em a priori} guaranteed but finds support in existing hybrid-kinetic simulations~\cite{st2018fluctuation}), the parallel rate of strain is largest at the parallel viscous scale $\ell_{\nu \parallel} \sim L\, \text{Re}_\parallel^{-3/4}$.
The parallel rate of strain, and thus the growth rate of magnetic fields, can then be expressed as
\begin{equation}
    \gamma \equiv \frac{{\rm d}\ln B}{{\rm d}t}  \simeq \ROS \sim \frac{U_{\rm rms}}{L} {\rm Re}_\parallel^{1/2}.
    \label{eq:db_dt}
\end{equation}
In what follows, we adopt (and adapt) arguments made by~\citet{schekochihin2006turbulence} for how the dependence of ${\rm Re}_{\parallel}$ on the magnetic-field strength through the action of these kinetic instabilities might lead to an accelerating dynamo and explosive growth of magnetic fields (phase~III).

As the flows stretch and amplify the magnetic fields, the mirror and firehose instabilities are triggered and their associated fluctuations grow.
In the early phase of the dynamo, when magnetic fields are sufficiently weak, the scattering rate needed to regulate $\Delta$ to within the firehose and mirror thresholds is larger than the ion Larmor frequency, that is, $|\ROS|/\Delta \gg \Omega_i$. Such a collisionality cannot be realized, i.e., the kinetic instabilities cannot scatter particles so fast that the plasma de-magnetizes. 
In this regime, it is reasonable to believe that the particle scattering rate is controlled by the growth rate of the mirror and firehose instabilities, both of which are proportional to the Larmor frequency times some power of the pressure anisotropy. We therefore write $\nu_{\rm eff} \propto B^\alpha$ with $\alpha$ being positive~\cite{schekochihin2006turbulence,melville2016pressure}.
Then, the dynamo growth rate ${\rm d}\ln B/{\rm d}t \propto 
{\rm Re}_{\parallel}^{1/2} \propto \nu_{\rm eff}^{1/2} \propto B^{\alpha/2}$ increases with increasing field strength, resulting in an  explosive growth of magnetic energy (characterized by a finite-time-singularity):
\begin{equation}
    \beta^{-1} = \beta_{\rm seed}^{-1}\left[1-\frac{\alpha}{2}\frac{U_{\rm rms}}{L}{\rm Re}_{\parallel 0}^{1/2}(t-t_{\rm seed})\right]^{-2/\alpha},
    \label{eq:beta_explosive}
\end{equation}
where $t_{\rm seed}\sim L/U_{\rm rms}$ indicates the moment of time at the beginning of this explosive phase. Here ${\rm Re}_{\parallel 0}$ is the parallel Reynolds number at this time, which is expected to be provided by phase mixing, collisions between particles, and/or the weak scattering off Weibel fluctuations, with an estimated value of~${\rm Re}_{\parallel 0} \sim \mathcal{O}(1)$.
Given the form of Eq.~\eqref{eq:beta_explosive}, the actual value of $\alpha$ does not affect the main feature of this phase. 
At the early time of this stage $(t-t_{\rm seed}\ll L/U_{\rm rms})$, the magnetic energy grows linearly $\beta^{-1} = \beta_{\rm seed}^{-1} [1+(U_{\rm rms}/L){\rm Re}_{\parallel0}^{1/2}(t-t_{\rm seed})]$.
The dynamo amplification starts to accelerate as it evolves with time and the explosive phase ends at around $(\alpha/2)(U_{\rm rms}/L){\rm Re}_{\parallel 0}^{1/2}(t-t_{\rm seed}) \sim 1$.
Because both $\alpha$ and ${\rm Re}_{\parallel0}$ are order-unity numbers, this phase III is expected to last for $(t-t_{\rm seed}) \sim L/U_{\rm rms}$, roughly one more flow-crossing time. 
This phase of accelerating dynamo growth, although being short, is critical to the overall amplification of magnetic fields because it allows the ${\rm Re}_{\parallel}$, and thus the dynamo growth rate, to increase to a comparatively large value.
Note that this phase has not yet been clearly realized in kinetic simulations, though finds some support in dedicated studies of the firehose and mirror~\citep{melville2016pressure}.

\subsection{Pressure-anisotropy-instability regulated dynamo.}
\label{sec:theory_dynamo}
As the nonlinear mirror and firehose fluctuations continue to grow alongside the dynamo field, the scattering of particles eventually becomes efficient enough to regulate the plasma anisotropy to values comparable to the instabilities' thresholds, {\em viz.}~$|\Delta| \sim 1/\beta$. We may then use Eq.~\eqref{eq:delta_brag} to estimate the value of $\nu_{\rm eff}$ required for this to occur, namely, $\nu_{\rm eff} \approx |\ROS|\beta$. (Such a collisionality has been measured directly in the later stages of the plasma dynamo using hybrid-kinetic simulations; \citealt{st2018fluctuation}.) 
The transition to this phase occurs when the increasing Larmor frequency (as the magnetic field grows) becomes comparable to this required collisionality, {\em viz.}~$\Omega_i \sim |\ROS|\beta$, which is equivalent to the requirement that \cite{st2020fluctuation}
\begin{equation}
    \beta^{-1}_{\rm ani} \sim (L/d_i)^{-2/5}M^{6/5}.
    \label{eq:beta_ani}
\end{equation}
Once this level of magnetic energy is reached, the hypothesized phase of explosive growth should end and the dynamo will start to be regulated by the pressure-anisotropy-instability (phase~IV).
Under typical ICM conditions, this value of $\beta^{-1}_{\rm ani}$ corresponds to ${\sim}{\rm nG}$ fields. That is, the magnetic fields are expected to be amplified to have a similar amount of energy as the Weibel fields before they coalescenced and decayed, but presumably with a much larger coherence scale than the Weibel fields. This phase of dynamo starts with a large effective ${\rm Re}_{\parallel}$ and thus the strength of the magnetic fields increases significantly. As derived and numerically confirmed in~\citet{melville2016pressure,st2018fluctuation}, the expression for the collisionality $\nu_{\rm eff} \approx 3|\ROS|\beta$ when the anisotropy is regulated suggests that 
\begin{equation}
    \text{Re}_\parallel \sim \nu_{\rm eff} U_{\rm rms}L/v_{\rm th}^2  \sim M^4 \beta^2.
    \label{eq:re}
\end{equation}
That is, as magnetic fields are amplified by this pressure-anisotropy-instability-regulated dynamo, the parallel Reynolds number, and thus the parallel rate of strain, keeps decreasing. [Note that Eq.~\eqref{eq:re} implies a parallel viscous scale that is commensurate with the scale on which the flow is Alfv\'enic, and so the dynamo is intrinsically nonlinear during this phase.]
The amplification of the magnetic field thus gradually slows down.
This phase eventually ends either when the dynamo saturates with approximate equipartition between the mean kinetic and magnetic energies, {\em viz.}~$\beta_{\rm sat}^{-1} \sim M^2$ and $\text{Re}_\parallel \sim 1$, or when the effective collisionality drops below the background Coulomb collisionality and the plasma is no longer kinetically unstable, {\em viz.}~$\beta_{\rm sat}^{-1} \lesssim M^{3/2} (\lambda_{\rm mfp,c}/L)^{1/2}$. In the latter case, the magnetic fields would continue to be amplified to equipartition. Coincidentally or not, the ICM seems to reside near the boundary between these two cases, with $\beta^{-1}\sim 10^{-3}$--$10^{-2}$, $M\sim 0.1$, and $\lambda_{\rm mfp,c}/L \sim 10^{-2}$--$10^{-1}$, and thus the anomalous scattering from the putative firehose/mirror instabilities is comparable to Coulomb scattering.

Though the aforementioned processes (except for the explosive phase) have been investigated independently with specific set-ups, how they transition from one to the other and how they collectively shape the collisionless turbulent dynamo is still unclear.
In Section~\ref{sec:numerical}, we present a numerical study that aims to include self-consistently all the relevant physical processes (for the case of a pair plasma). Unfortunately, quite a large scale separation $L/d_i$ is required to satisfy both $\beta_{\rm seed}^{-1} \ll \beta_{\rm ani}^{-1} \ll \beta_{\rm sat}^{-1}$ and $\beta^{-1}_{\rm seed} \ll \beta^{-1}_{\rm w}$, and distinguish between all of the hypothesized four phases of evolution. 
Using $M \sim 0.1$ and assuming these critical values of $\beta^{-1}$ are separated by at least a factor of 10, we require that $L/d_i \gtrsim 10^5$. 
This requirement vastly exceeds that which can be achieved with today's computational resources [the cost of a simulation scales as $(L/d_i)^4$], and the simulation discussed in the next section is only able to provide qualitative evidence for many of these theoretical predictions.

\section{Numerical experiments}
\label{sec:numerical}

\subsection{Numerical methods}
We perform fully kinetic, particle-in-cell (PIC) simulations to study the plasma dynamo with the code \textit{Zeltron}~\cite{Cerutti2013}.
Because of the high computational cost inherent to this problem, our simulations are performed with an electron-positron plasma, in which the skin depths of both species are identical, as are their Larmor-radius scales.
The system is initialized with a spatially uniform, isotropic, unmagnetized, Maxwell--J\"{u}ttner plasma of sub-relativistic temperature $T_{e0} \equiv\theta_e m c^2=1/16$.
It is continuously subjected to a random, time-correlated external volumetric mechanical force, $\bb{F}_{\rm ext}$, applied at the largest scales of the domain~\citep{zhdankin_2021} (details in~SM). 
The force $\bb{F}_{\rm ext}$ contains six solenoidal modes with time-dependent random phases and is designed to drive incompressible flows.
Optically thin external inverse Compton (IC) radiative cooling (parameters described in~SM) is included to achieve a steady temperature close to $T_{{\rm e}0}$ and to suppress nonthermal particle acceleration by cooling mainly at the high-energy tails of the plasma distributions. 
The IC radiation is isotropic and thus is not expected to affect the properties of the plasma dynamo.
The simulation is performed in a 3D periodic cubic box and the separation between the domain scale ($L$) and the plasma skin depth ($d_e$) is $L/d_e=378$, so that the total number of cells is~$1512^3$. 
We use 32 particles per cell (PPC) (16 per species) for the simulation and so around 100 billion particles in total. 
Simulations with PPC ranging from 32 to 128 show the same results.
The grid spacing is uniform with ${\rm d} x ( ={\rm d} y = {\rm d} z) = \lambda_{{\rm D}e} = d_e/4$, where $\lambda_{{\rm D}e}$ is the (initial) Debye length and $d_e/\lambda_{{\rm D}e}=\sqrt{1/\theta_e}=4$.

\subsection{Numerical results}
\label{sec:results}
The overall evolution of the system in our simulation can be divided into four qualitatively different stages: the linear Weibel stage ($tU_{\rm rms}/L \lesssim \tau_{\rm w}$), the Weibel-filament coalescing stage ($\tau_{\rm w} < tU_{\rm rms}/L \lesssim 1$), the exponential dynamo stage ($1 < tU_{\rm rms}/L \lesssim 3$), and then slow amplification until saturation ($3 \lesssim tU_{\rm rms}/L \lesssim 7$).

\begin{figure}[h!]
    \centering
    \includegraphics[width=0.45\textwidth]{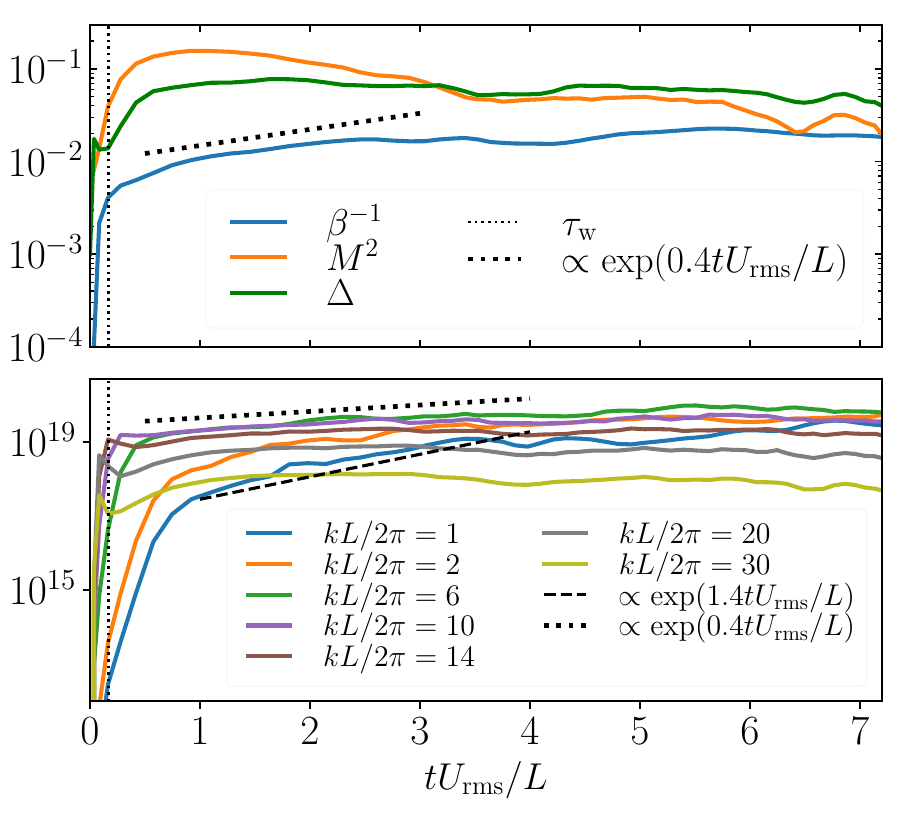}
    \caption{Top: time evolution of $M^2$, $\Delta$, and $\beta^{-1}$. Bottom: time evolution of magnetic energy density at various wavenumbers. The vertical dotted line indicates the time $\tau_{\rm w}$ when the Weibel instability reaches its largest growth rate. }
    \label{fig:M_delta_beta}
\end{figure}

\begin{figure*}[t!]
    \includegraphics[width=0.95\textwidth]{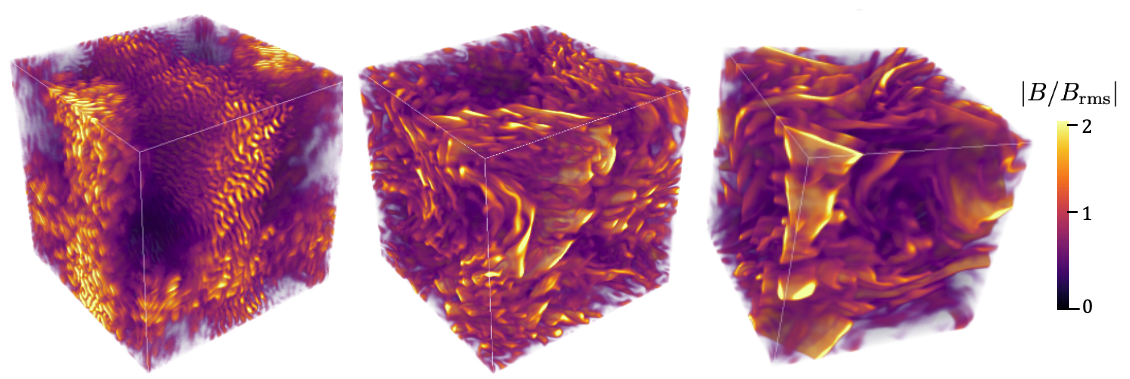}
    \caption{Visualization of (normalized) magnetic fields' magnitude at peak Weibel growth ($tU_{\rm rms}/L=\tau_{\rm w}$; left), after one large-scale turnover time ($t=L/U_{\rm rms}$; middle), and in the saturated state of the dynamo (right).}
    \label{fig:visuals}
\end{figure*}

\paragraph{Generation and evolution of seed magnetic fields.}
The energetics of the system are described by the time evolution of $\beta^{-1}$, $M^2$, and $\Delta$ (defined as described in Section~\ref{sec:theory} but instead using the electron mass and electron thermal speed), shown in the top panel of Figure~\ref{fig:M_delta_beta}. 
As the plasma is stirred by the applied force, bulk motions are established and $M^2$ increases. 
The phase mixing of the inhomogeneous flows by the thermal motion of the particles leads to the development of pressure anisotropy~$\Delta$.
In an unmagnetized plasma, this anisotropy triggers the Weibel instability~\cite{weibel1959,burton1959} and generates the seed magnetic fields, as indicated by the rapid increase of~$\beta^{-1}$. 
The Weibel instability reaches its largest growth rate (measured using the growth rate of the volume-averaged magnetic field strength) at a time that we denote $\tau_{\rm w}$ (${\approx}0.16 L/U_{\rm rms}$; vertical dotted line), at which point the morphology of the magnetic field is shown in the left panel of Figure~\ref{fig:visuals}.  
Clear filamentary structures of the Weibel fields on ${\sim}d_e$ scales emerge from the initial random noise and occupy an appreciable fraction of the volume of the domain.

\begin{figure}[h!]
    \centering
    \includegraphics[width=0.45\textwidth]{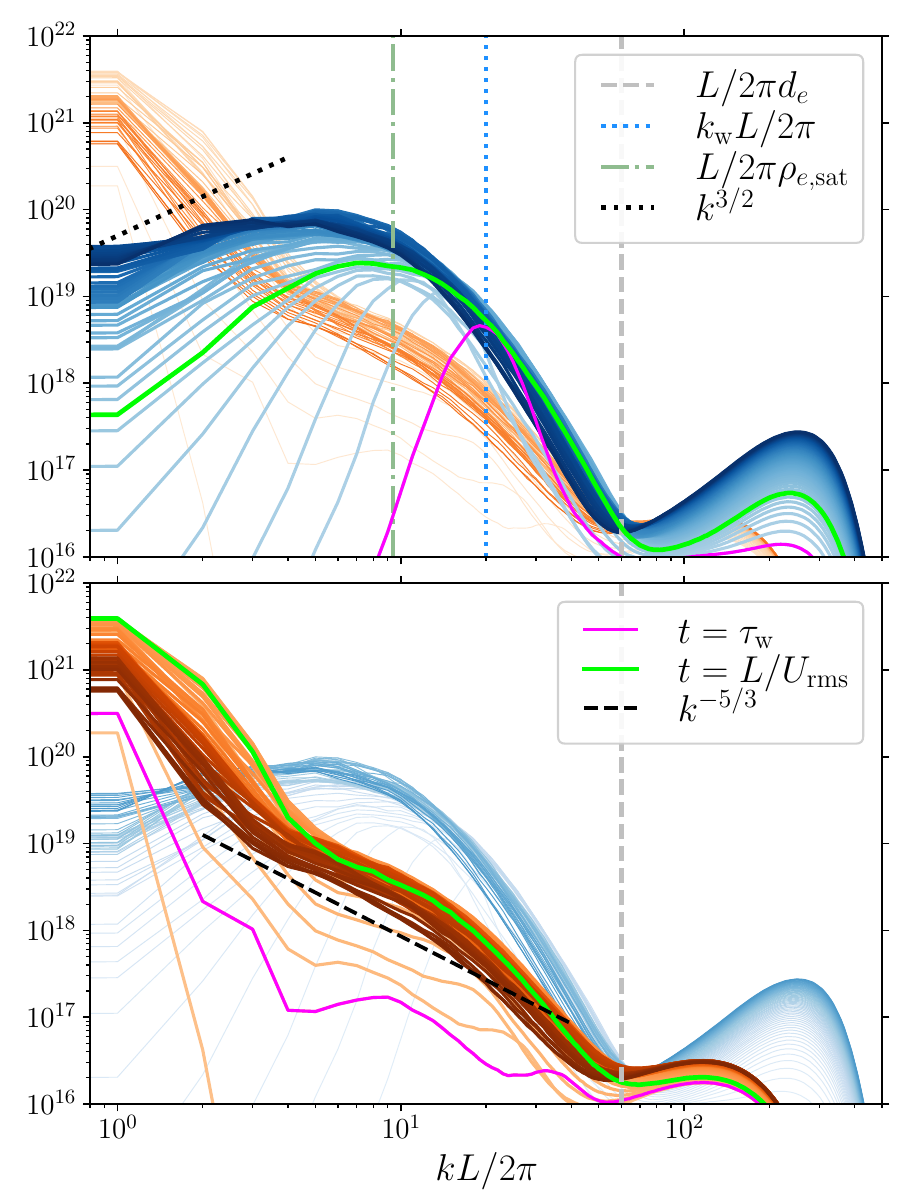}
    \caption{Magnetic (blue) and bulk kinetic (orange) energy spectra at various times, with earlier times corresponding to lines with greater transparency. The top (bottom) panel highlights the magnetic (kinetic) spectrum, with the other  shown in the background for reference. 
    The spectra at $tU_{\rm rms}/L=\tau_{\rm w}$ ($L/U_{\rm rms}$) are highlighted in magenta (green). 
    The magnetic (kinetic) spectrum at $t=L/U_{\rm rms}$ is highlighted with a black solid (dashed) curve.
    The silver, blue, and green vertical lines indicate the scales of the electron skin depth, the Weibel filaments, and the Larmor radius of particles after dynamo saturation, respectively.}
    \label{fig:spectra}
\end{figure}

The time evolution (over the whole simulation) of magnetic and bulk kinetic energy spectra [integrated on spherical shells in wavenumber ($\bb{k}$) space], $\mathcal{M}(k)\equiv (1/4\pi) \int {\rm d}\Omega_k \, k^2 |\bb{B}(\bb{k})|^2/8\pi$  (blue lines) and $\mathcal{K}(k)\equiv (1/4\pi) \int {\rm d}\Omega_k \, k^2 |\bb{u}(\bb{k})|^2\braket{\rho} /2$ (orange lines), is shown in Figure~\ref{fig:spectra}, with earlier times corresponding to curves with greater transparency.
The magnetic (kinetic) spectra with blue (orange) curves are highlighted in the top (bottom) panel, with the other one shown in the background for reference.
At early times (around $\tau_{\rm w}$), the bulk kinetic energy concentrates at the system scale, with little energy cascading to the smaller scales. 
This is consistent with the expectation that a collisionless, unmagnetized plasma is very viscous with the effective ${\rm Re}_{ \parallel} \sim 1$ due to unimpeded phase mixing.
The magnetic energy spectrum at $\tau_{\rm w}$ (highlighted with the magenta curve) peaks at $k_{\rm w} L/2\pi \approx 20$, i.e., a wavenumber $k_{\rm w} \approx 0.3/d_e$.

The scaling dependence of the saturated amplitude and length scale of the Weibel seed fields on the rate of strain of the flow ($U_{\rm rms}/L$) and the scale separation~($L/d_e$), as determined by~\citet{zhou2022spontaneous}, is given by Eq.~\eqref{eq:beta_weibel}.
With the limited scale separation in our simulation, we expect the Weibel instability to produce seed fields with energy $\beta^{-1}_{\rm w} \simeq 10^{-2}$ and wavenumber $k_{\rm w}d_e \simeq 0.2$. This is roughly consistent with our measured $\beta^{-1} \approx 3\times 10^{-3}$ and $k_{\rm w}d_e \approx 0.3$ at $\tau_{\rm w}$ in Figure~\ref{fig:M_delta_beta}.
For comparison, using a value of $L/d_e \sim 10^{16}$ characteristic of the bulk ICM in Eq.~\eqref{eq:beta_weibel} predicts $\beta_{\rm w}^{-1} \sim 10^{-8}$. 

\begin{figure}[!]
    \centering
    \includegraphics[width=0.45\textwidth]{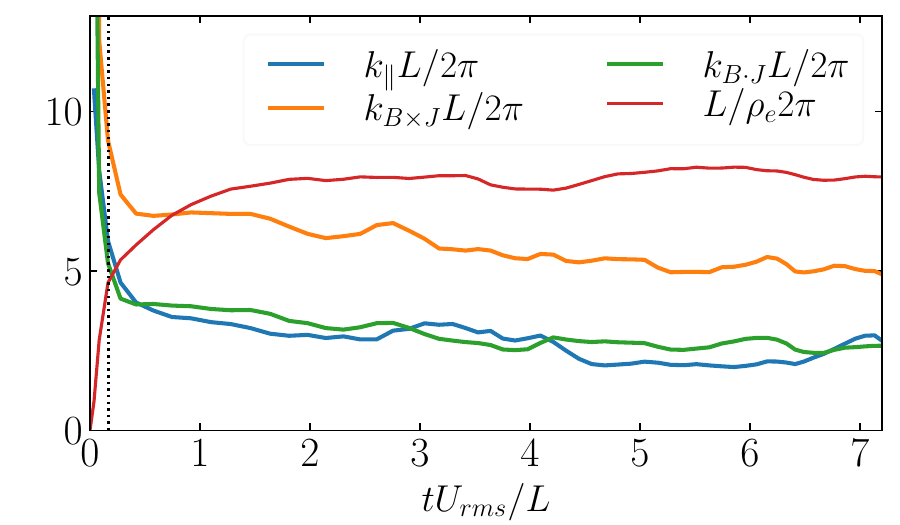}
    \includegraphics[width=0.4\textwidth]{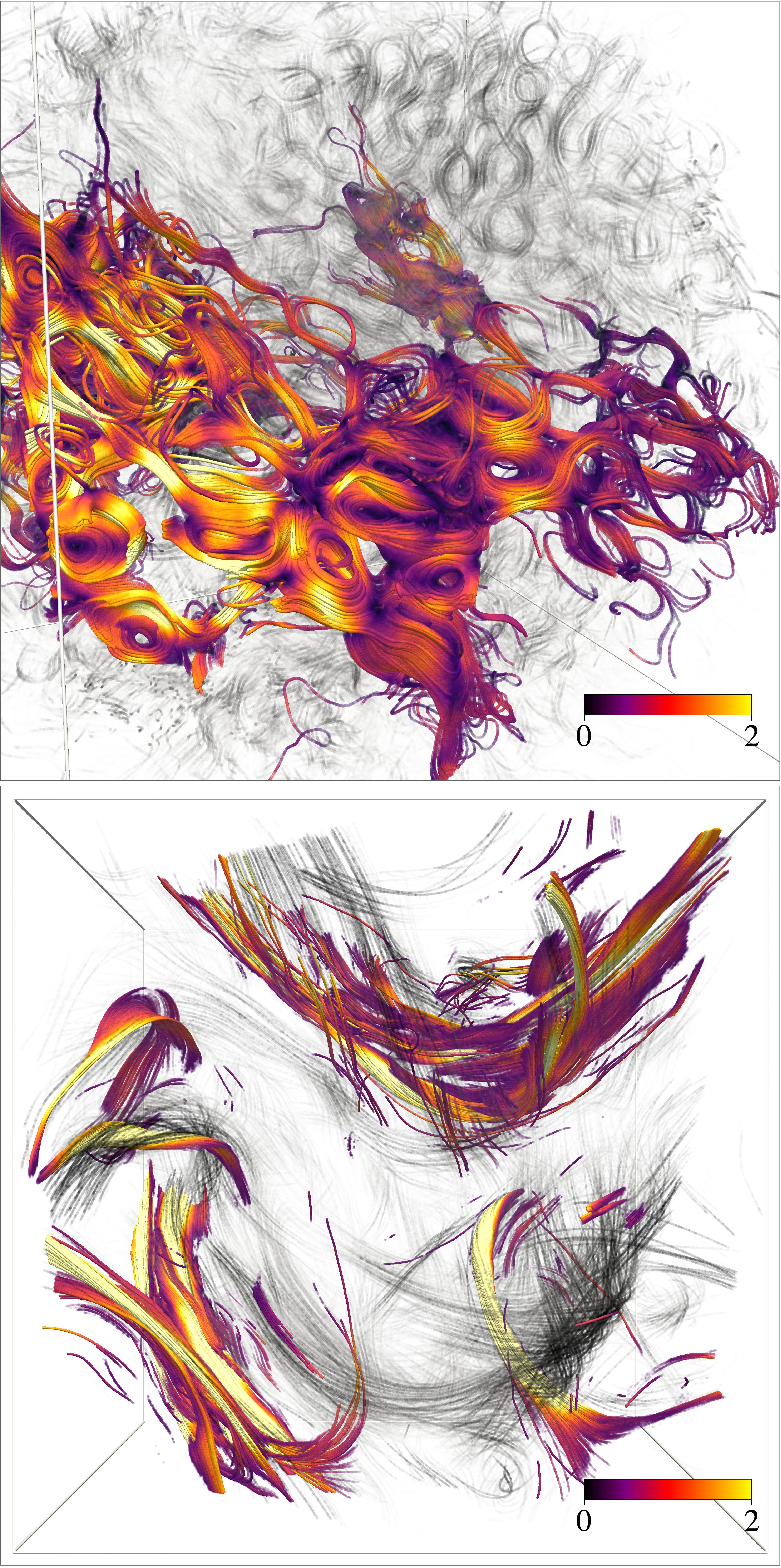}
    \caption{Morphology of the magnetic field. Top: time evolution of characteristic wavenumbers describing the magnetic field and the averaged Larmor radius. The vertical dotted line indicates the time $\tau_{\rm w}$. Magnetic-field lines at $t=0.3 L/U_{\rm rms}$ (middle) and at saturation (bottom) are shown. The colored field lines are chosen from regions with strong magnetic field. The grey shaded field lines are randomly sampled through the domain. The color corresponds to the normalized field strength, $|B/B_{\rm rms}|$. }
\label{fig:wavenumbers}
\end{figure}

After $\tau_{\rm w}$, the system enters the second stage in which the nonlinear effects of the Weibel instability become important.
The peak of the magnetic spectrum shifts to lower $k$ while continuing to grow in amplitude. 
Two effects are responsible for this shift.
The first is due to the low-$k$ Weibel modes. While the high-$k$, fastest-growing Weibel modes become nonlinear and stop growing exponentially, the initially subdominant longer-wavelength Weibel modes are still in the linear stage and continue to grow exponentially, albeit with a lower linear growth rates, and thus start to overtake the high-$k$ modes.
The second, as described in Sec.~\ref{sec:theory_coalescence}, the Weibel filaments, after forming, start to coalesce  with one another via reconnection~\cite{zhou2020multi} before being affected by the flow shear on the time scale ${\sim} L/U_{\rm rms}$.
Evidence for coalescence can be seen in the middle panel of Figure~\ref{fig:wavenumbers}, which plots representative field lines chosen from regions with strong magnetic fields at $t =0.3 L/U_{\rm rms}$.
Through this process, the coherence length scale of the magnetic fields grows and the magnetic energy dissipates rapidly. 
This can be seen more clearly in the time evolution of magnetic energy densities at different~$k$, shown in the bottom panel of Figure~\ref{fig:M_delta_beta}. 
Initially, the fastest growing modes occur at around $kL/2\pi \approx 20$ (i.e., at wavenumber~$k_{\rm w}$).
These modes saturate quickly (at around~$\tau_{\rm w}$), and are immediately followed by a sudden drop of their energy densities due to filament coalescence, coinciding with a jump in energy density at smaller wavenumbers by a factor of $\sqrt{2}$ (consistent with flux conservation~\cite{zhou2020multi}).

With sufficient length scale separation $L/d_e$ (and hence~$k_{\rm w} L$) and time scale separation between $\tau_{\rm w}$ and~$L/U_{\rm rms}$, we anticipate a prolonged phase of successive coalescence of Weibel filaments. 
The decay of magnetic energy and the growth of coherence length scale during such a phase is expected to follow the scaling laws given by Eq.~\eqref{eq:inverse_law}. 
In our simulation, only a couple of generations of coalescence are allowed before being interrupted by the flow shear and this predicted phase in which $\beta^{-1}$ decreases is not well captured.

\paragraph{Establishment of plasma dynamo.}
After the phase of Weibel growth and magnetization of the plasma, the system is continuously driven by the external force towards reaching a statistical steady state. 
As shown in the top panel of Figure~\ref{fig:M_delta_beta}, both $M^2$ and $\Delta$ continue to increase and attain steady values by $t \approx L/U_{\rm rms}$.
On this flow-crossing time scale, the large-scale flows are fully developed and start to stretch and deform the seed fields. 
The (coalesced) Weibel seed fields reorganize, aligning with the large-scale flows, as shown in the middle panel of Figure~\ref{fig:visuals}.
The magnetic and kinetic energy spectra at $t =L/U_{\rm rms}$ are highlighted in Figure~\ref{fig:spectra} with green curves. 
The kinetic spectrum rapidly extends towards smaller scales, with a knee forming at $kL/2\pi \approx 5$, where the kinetic and magnetic energy are comparable. 
Below the knee, a shallower kinetic spectrum is formed, where a $k^{-5/3}$ line is included as a visual reference.
The broadening of the kinetic spectrum indicates that some energy is cascading to smaller scales and suggests the formation of a viscous scale slightly smaller than the outer scale.
The magnetic spectrum is also broadened due to the combined effect of filament coalescence and rearrangement, the growth of low-$k$ Weibel modes, and the stretching by the flow.
As a consequence of limited scale separation, when the bulk flows are well established, the magnetic energy density is already larger than the kinetic one across a wide range of scales (at $kL/2\pi \gtrsim 5$).
This leaves a narrow range for the action of plasma dynamo, because the stretching and folding of magnetic-field lines can only happen at scales where the flows are energetically dominant.

The geometry of the magnetic field can be characterized using the wavenumbers $k_\parallel \equiv \left( \braket{|\bb{B \cdot \nabla B}|^2}/\braket{B^4} \right)^{1/2}$, $k_{B \times J} \equiv \left( \braket{|\bb{B \times J}|^2}/\braket{B^4} \right)^{1/2}$, and $k_{{B \cdot J}} \equiv \left( \braket{|\bb{B \cdot J}|^2}/\braket{B^4} \right)^{1/2}$, which quantify the variation of the magnetic field along itself, in the direction of the field reversal, and in the cross direction, respectively~\cite{schekochihin2004simulations}. 
If the magnetic field possesses a folded structure, its length, thickness, and width can be represented by $\ell \sim 1/k_\parallel$, $\lambda \sim 1/k_{B \times J}$, and $\xi \sim 1/k_{{B \cdot J}}$, respectively.
The time evolution of $k_\parallel$, $k_{B \times J}$, and $k_{B \cdot J}$ is shown in the top panel of Figure~\ref{fig:wavenumbers}, together with the domain-averaged normalized inverse particles' Larmor radius $(L/2\pi)/\rho_e$.
The three wavenumbers initially have large values set by the Weibel filaments, and drop rapidly from the beginning of the simulation to $t \approx 0.4 L/U_{\rm rms}$ due to the rapid disentanglement of the helical field lines by the turbulent flows, visualized in the middle panel of Figure~\ref{fig:wavenumbers}. 
Starting at $t \approx 0.4 L/U_{\rm rms}$, the wavenumbers decrease slowly to become nearly constant, with the ordering $k_\parallel \approx k_{B \cdot J} \ll  k_{B \times J}$, indicative of a folded sheet-like structure.
The inverse Larmor radius $(L/2\pi)/\rho_e$ increases first rapidly up until $\tau_{\rm w}$ and then slower between $\tau_{\rm w}$ and~$L/U_{\rm rms}$. 
As the growth of magnetic energy drastically slows down after $L/U_{\rm rms}$, $(L/2\pi)/\rho_e$ also reaches an approximate saturation.
At $t\approx L/U_{\rm rms}$, $k_{B \times J} \approx 1/\rho_e$ is observed, suggesting that the thickness of the folds is comparable to the Larmor radii of particles.
This is consistent with the argument that the length scale for the magnetic field reversal cannot become shorter than the local Larmor radius as a result of the stretching and folding of the flows, as the field lines are not frozen into the plasma below the Larmor radius. 

\paragraph{Inductive dynamo amplification.}
As the turbulent flows are established and the seed field becomes coupled to the flow, it becomes possible for the flows to stretch and fold the magnetic field lines.
The statistical outcome of this process is an inductive amplification of the magnetic field, known as the plasma dynamo, which leads to another period of exponential growth of magnetic energy at $L/U_{\rm rms} \lesssim t \lesssim 3L/U_{\rm rms}$ in our simulation, as evidenced in Figure~\ref{fig:M_delta_beta}.
The overall growth of magnetic energy can be fit with $\beta^{-1} \propto \exp(0.4U_{\rm rms}t/L)$, corresponding to a magnetic growth rate $\gamma \equiv {\rm d} \ln B/{\rm d}t \approx 0.2 U_{\rm rms}/L$.
This growth rate is tied to the macroscopic eddy-turn-over rate and is thus much slower than the rapid growth on the kinetic time scale during the Weibel stage.
The dynamo growth rate being comparable to the flow-crossing rate suggests that the inductive amplification is given by the flow on scales not far removed from the domain scale, and is another manifestation of the lack of scale separation in our simulation:
it is consistent with the fact that, at $t \approx L/U_{\rm rms}$, the seed field is so strong that only at scales comparable to the domain size are bulk flows energetically dominant (see Figure~\ref{fig:spectra}).
The strong magnetic field exerts a back reaction on the plasma flow through the Lorentz force, causing the dynamo to start in an already nonlinear regime, skipping the kinematic phase.
As shown in the bottom panel of Figure~\ref{fig:M_delta_beta}, at larger scales where the strength of the seed field is smaller, the dynamo growth rate at $L/U_{\rm rms}\lesssim t \lesssim 3L/U_{\rm rms}$  is higher (e.g., $\gamma \approx 0.7 U_{\rm rms}/L$ for $kL/2\pi=1$) due to the weaker back reaction of the field on the flow.  
This leads to a further broadening of the magnetic spectrum and accumulation of energy at the system scale (Figure~\ref{fig:spectra}, top panel). 
The magnetic spectrum at large scales is flatter than $k^{3/2}$ (shown with the black dotted line), which is expected for the kinematic dynamo~\cite{Kazantsev1968}).
This is consistent with the observation that the dynamo starts in a nonlinear regime.
For $t\gtrsim 3L/U_{\rm rms}$, as the magnetic energy becomes larger and approaches the kinetic energy, the growth of the magnetic field slows down further.

The time evolution of the characteristic wavenumbers and Larmor radius shown in Figure~\ref{fig:wavenumbers} is consistent with the amplification of the magnetic field and the broadening of the magnetic spectrum towards larger scales.
As the magnetic field grows in strength, the domain-averaged Larmor radius decreases [$(L/2\pi)/\rho_e$ increases] and becomes smaller than the length scale of the magnetic field in all dimensions. 
All three wavenumbers slowly decrease, with $k_\parallel$ and $k_{B \cdot J}$ approaching $kL/2\pi \approx 2$ to 3 at late times. 
This suggests the formation of magnetic folds with sizes comparable to the system scale.
Such large-scale folded sheets are clearly seen in the right panel of Figure~\ref{fig:visuals}.

The dynamo eventually saturates by the end of the simulation when approximate equipartition between kinetic and magnetic energy is reached, $M^2 \beta \sim 1$ (Figure~\ref{fig:M_delta_beta}, top panel).
As shown in Figure~\ref{fig:spectra}, the peak of the magnetic energy spectrum continues to shift to larger scales until saturation (with the energy density at small $k$ increasing and that at large $k$ decreasing).
Although an overall equipartition is reached, the scale-by-scale equipartition does not exist in any scale range of the spectra. 
The kinetic energy dominates at small $k$ whereas the magnetic energy dominates at large~$k$. 
Starting from $t \sim L/U_{\rm rms}$, the local (in $k$-space) equipartition [$\mathcal{M}(k) \sim \mathcal{K}(k)$] first occurs at around $kL/2\pi \approx 5$ and then slowly shifts to smaller~$k$. The structure of the turbulence and the energy cascade in this turbulent dynamo remain to be investigated. 

\begin{figure*}[t]
    \includegraphics[width=0.99\textwidth]{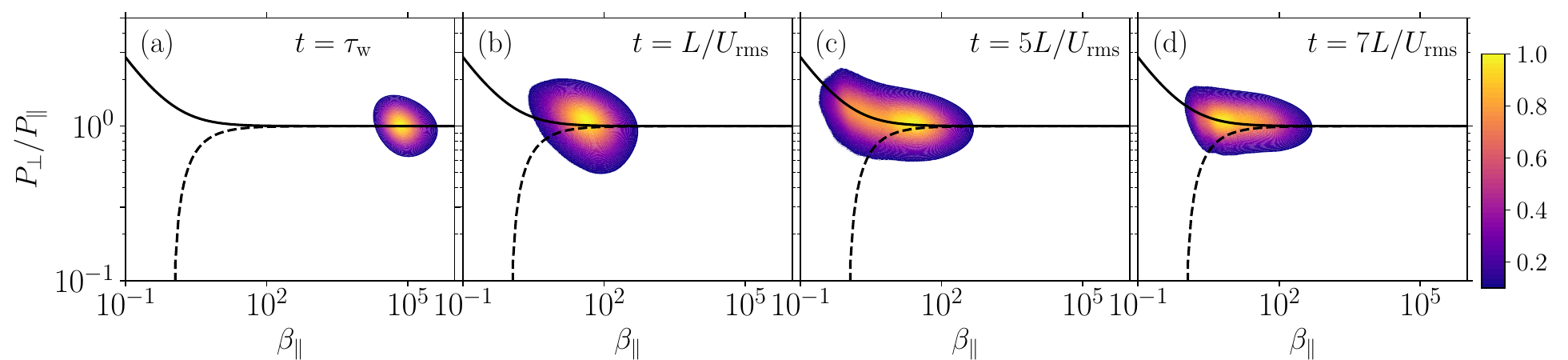}
    \caption{Normalized probability density distribution of pressure anisotropy and $\beta_{\parallel}$ at various moments of time. The solid (dashed) curve represents the threshold for the mirror (firehose) instability.}
    \label{fig:brazil_plot}
\end{figure*}

\begin{figure}[h!]
    \centering
    \includegraphics[width=0.45\textwidth]{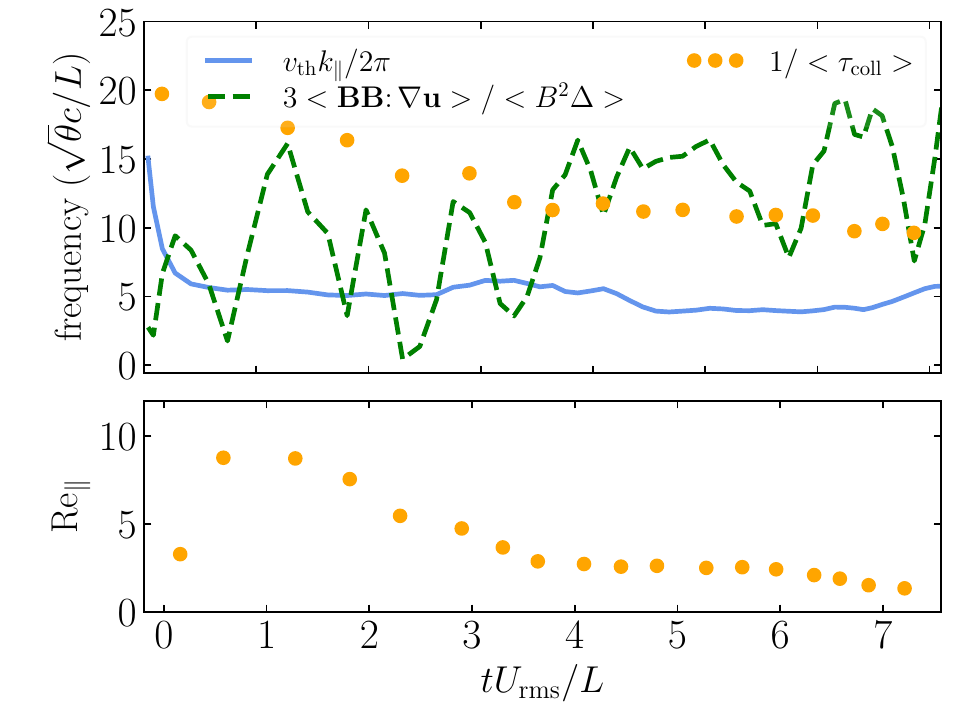}
    \caption{Top: time evolution of normalized (to $\sqrt{\theta}c/L$) effective collisionality $\nu_{\rm eff} =1/\braket{\tau_{\rm coll}}$ (orange), compared to a parallel streaming frequency $v_{\rm th}k_\parallel/2\pi$ (blue), and a ``Braginskii'' collision frequency assuming Eq.~\eqref{eq:delta_brag} (green). Bottom: effective parallel Reynolds number implied by~$\nu_{\rm eff}$. }
    \label{fig:nu_eff}
\end{figure}

\paragraph{Instability-regulated turbulence.}
In this fully collisionless turbulent dynamo, important parameters such as the viscosity and resistivity are self-consistently determined by the interaction between the plasma and micro-scale fluctuations. 
The large-scale motions in a collisionless plasma push it out of local thermodynamic equilibrium, and various kinetic instabilities can become unstable due to the developed pressure anisotropy.  
After the triggered Weibel instability magnetizes the plasma, the Larmor-scale instabilities, namely the mirror and firehose instabilities, become unstable when $|P_\perp - P_\parallel| \gtrsim B^2/4\pi$.  

The presence of these kinetic instabilities in our simulation is indicated in the probability density distribution of $P_\perp/P_\parallel$ and $\beta_\parallel$ shown in Figure~\ref{fig:brazil_plot}.
During the Weibel phase, the kinetic-scale filamentary seed field naturally causes a short mean-free-path of the particles, as the largest distance that a particle can travel before  pitch-angle scattering is not longer than the length of the Weibel filaments.
The scattering off the Weibel seed field confines the pressure anisotropy at small values, i.e., $P_\perp \approx P_\parallel$ [panel (a)].
The saturated seed field is then organized into folded-sheet structures (at $t \approx L/U_{\rm rms}$) with coherence scales much larger than the Weibel field.
As described in Section~\ref{sec:theory_dynamo}, due to the near conservation of~$\mu$, local changes in the magnetic-field strength trigger the mirror and firehose instabilities.
The Larmor-scale fluctuations generated by these instabilities start to grow on top of the folded magnetic field and scatter the particles, limiting the growth of pressure anisotropy [panel~(b)]. 
The growing fluctuations eventually become strong enough to pin the pressure anisotropy around the mirror instability's threshold [e.g., at $t=5L/U_{\rm rms}$ shown in panel~(c)].
Because the strength of the magnetic field is overall growing, $P_\perp>P_\parallel$ is expected to populate the domain and the mirror instability is dominant. 
After the saturation of the dynamo, the regions where the field is amplified should balance those where the field is diminished, and therefore, both mirror and firehose instabilities are active, regulating the pressure anisotropy in between their thresholds [panel~(d)].

The scattering of particles off the mirror and firehose fluctuations provides an effective collisionality, $\nu_{\rm eff}$, which can be quantified by a pitch-angle scattering rate and measured using the trajectories of tracked particles.
One way of quantifying the scattering rate is to study the time evolution of particles' magnetic moments~$\mu$, and compute the histogram of the collision time~$\tau_{\rm coll}$, defined as the time interval for $\mu$ of each particle to change by a factor~${e}$. 
The characteristic collision time $\braket{\tau_{\rm coll}}$ is obtained by fitting the histogram with an exponential function (details see SM), and is the average time scale on which the conservation of $\mu$ is violated.
The effective collisionality thus obtained, $\nu_{\rm eff} = 1/\braket{\tau_{\rm coll}}$, is shown as a function of time in the top panel of Figure~\ref{fig:nu_eff}.
In the same panel, the time evolution of $v_{\rm th}k_\parallel/2\pi$ and $3\braket{\ROS}/\braket{B^2 \Delta}$ are shown for comparison.
The former represents the scattering rate assuming that the particles' mean free paths are comparable to the length of the magnetic folds, and the latter is based on the assumption that the anisotropy evolves following Eq.~\eqref{eq:delta_brag}, which yields a Braginskii-type scattering rate. 
At early times, $1/\braket{\tau_{\rm coll}} \approx v_{\rm th}k_\parallel/2\pi$ suggests that the particles stream parallel to the field lines until they are scattered at the end points of the magnetic structures.
At $t \simeq L/U_{\rm rms}$, the folded-sheet structures fully form and the Larmor-scale mirror and firehose fluctuations start to grow. 
From then on, $1/\braket{\tau_{\rm coll}} \approx 3\braket{\bb{BB:\nabla u}}/\braket{B^2 \Delta} \gg v_{\rm th}k_\parallel/2\pi$, suggesting that the collisionality is mainly caused by the particle scattering off micro-instabilities, which regulates the pressure anisotropy.
This is consistent with the fact that $\Delta$ reaches a steady state at $t \approx L/U_{\rm rms}$ (Figure~\ref{fig:M_delta_beta}, top panel).
 
An effective field-parallel Reynolds number is determined by the effective collisionality $\text{Re}_\parallel \equiv U_{\rm rms}L/(v_{\rm th}^2/\nu_{\rm eff})$, and its time evolution is shown in the bottom panel of Figure~\ref{fig:nu_eff}.
The $\text{Re}_\parallel$ first increases at $t<L/U_{\rm rms}$ as the flows develop, and then decreases until $\text{Re}_\parallel \approx 1$.
This is consistent with the expectation discussed in Section~\ref{sec:theory_dynamo} that, as the particle scattering off nonlinear Larmor-scale fluctuations becomes sufficient to regulate the pressure anisotropy, $\nu_{\rm eff}$ decreases as the magnetic energy increases.
This leads to a decreasing $\text{Re}_\parallel$ and an increasing parallel viscous scale~$\ell_{\nu \parallel}$, which could be the reason for the knee of the kinetic spectrum to shift slightly towards smaller wavenumbers, shown in the bottom panel of Figure~\ref{fig:spectra}.
The parallel rate of strain of the flow, and hence the dynamo growth rate, are expected to decrease, consistent with the progressively slower growth of magnetic energy before saturation (Figure~\ref{fig:M_delta_beta}). 
The detailed scaling dependence of $\text{Re}_\parallel$ [Eq.~\eqref{eq:re}], derived under the assumption of Kolmogorov scalings, is not expected to hold in our simulations due to the lack of scale separation and of an inertial range of the turbulence.   
 
\section{Conclusions and discussion}
\label{sec:discission}

In this work we present an analytical theory and a first-principles numerical demonstration of the generation, amplification, and sustenance of magnetic fields under the action of large-scale turbulent flows in a collisionless plasma.
In an initially unmagnetized Maxwellian plasma, the developed thermal pressure anisotropy resulting from the phase mixing of large-scale flows triggers the Weibel instability, which depletes the thermal free energy in the pressure anisotropy to produce a filamentary seed magnetic field at plasma-kinetic scales strong enough to magnetize the plasma. After a brief phase of filament coalescence, at around one flow-crossing time, the seed field becomes coupled to the large-scale flows.
The turbulent flows start to stretch and fold the magnetic field, producing a field-biased pressure anisotropy via the approximate conservation of the magnetic moments of magnetized particles. The mirror and firehose instabilities become unstable to this pressure anisotropy and generate Larmor-scale fluctuations.
The scattering of particles off of these fluctuations leads to an effective collisionality, which in return regulates the pressure anisotropy, controls the parallel rate of strain of the flow and thus the dynamo growth rate. The magnetic field is inductively amplified until it reaches approximate energy equipartition with the flow.
The length scale of the field is found to approach the system scale at saturation, suggesting that a collisionless fluctuation dynamo can produce a magnetic field that is coherent on scales comparable to the turbulence forcing scale.
Most importantly, our results provide a proof-of-principle demonstration that equipartition magnetic fields can be produced in an unmagnetized system by large-scale astrophysical flows without resorting to other magnetic seeding mechanisms. 

Despite these successes, the predictive ability of our numerical simulations is rather limited. In Section~\ref{sec:theory}, we introduced theoretical arguments and leveraged previous numerical simulations of the plasma dynamo to advance a four-phase evolutionary scenario for the {\em ab-initio} plasma dynamo (see Figure~\ref{fig:sketch}). In this scenario, magnetic fields are self-consistently seeded by the Weibel instability and amplified inductively by random bulk flows in a plasma whose viscosity is ultimately controlled by rapidly growing, microscale, mirror and firehose instabilities and therefore dependent upon the plasma~$\beta$. Some features of this scenario are borne out by our numerical simulations, but several others cannot be tested at this time because of the very large scale separation they require (namely, $L/d_i\gtrsim 10^5$, if not larger). In our simulation with $L/d_e =378$, processes such as the decrease of $\beta^{-1}$ during the predicted reconnection-controlled coalescence of the Weibel seed field and the kinematic (and potentially explosive) phase of the dynamo cannot be unambiguously identified.

Another unconstrained piece of physics concerns what sets the characteristic reversal scale of the amplified magnetic field, particularly during the ``kinematic'' linear phase in which the back reaction of the magnetic field on the flow (through the Lorentz force) is negligible and the magnetic energy resides at the smallest available scale. In the kinematic phase of the ${\rm Pm}\ge 1$ MHD dynamo, this scale is the resistive scale~\cite{Kazantsev1968,KulsrudAnderson1992,Schekochihin2002,GKS22}. In a collisionless dynamo, it is reasonable to expect this scale to instead be comparable to (or at least related to)~$\rho_e$, the scale below which the magnetic field is not ``frozen'' into the plasma and the flow cannot efficiently stretch and amplify the magnetic field. If true, then $1/k_{J \times B} \sim \rho_e \propto B^{-1}$ during the linear phase of the dynamo; as the magnetic field is amplified, the characteristic reversal scale of the field would then continuously shrink. This would be a unique feature of a fully kinetic dynamo.

The demonstrated self-consistent generation of near-equipartition magnetic fields under the action of large-scale turbulent flows has important implications for the origin of intracluster fields. 
The energy density of a ${\sim}\mu{\rm G}$ magnetic field in the ICM is usually comparable to that of the turbulent motions; for example, the energy density of a $10~\mu{\rm G}$ magnetic field approximately matches the kinetic energy density of a hydrogenic plasma with velocity dispersion ${\approx}164~{\rm km~s}^{-1}$ at a number density ${\approx}0.02~{\rm cm}^{-3}$, parameters measured in the ICM of Perseus~\citep{hitomi2016quiescent}).
This suggests that astrophysical turbulence may itself explain the observed ${\sim}1$--$10~\mu{\rm G}$ intracluster fields~\citep{KJZ22}.
Future numerical studies that achieve asymptotically large scale separation, or perhaps reduced models adopting accurate microphysical closures, are needed to further test this statement. 

Finally, we note that another important and oft-advocated origin of cosmic seed magnetic fields---a proto-galactic origin---is not considered in this work.
Scenarios based on a proto-galactic origin postulate that seed fields are generated (by various fluid or plasma-kinetic instabilities) through gravitational collapse or collision-related events occurring during structure formation and stellar evolution within galaxies, and are then injected into and diluted throughout the intergalactic medium by powerful galactic winds or jets~\cite{rees1968model,rees1987origin,furlanetto2001intergalactic}. 
The conjecture of a galactic origin is supported indirectly by observations of early enrichment of galaxy clusters by metals~\cite{mantz2020deep}, namely, {\it XMM-Newton} observations of a cluster at redshift $z \simeq 1.7$ with high metal enrichment (${\sim} 1/3$ Solar). 
If such galactic pollutants were accompanied by ${\sim}\mu{\rm G}$ galactic magnetic fields, it is possible that the seed fields in the ICM have a galactic origin. 
However, the feasibility of this seeding mechanism depends upon the efficiency with which such fields are dispersed and diluted throughout a turbulent, weakly collisional ICM, a process that remains to be studied.
This paper explores the scenario that magnetic fields can be originated from the gravitationally driven macroscopic flows in the ICM without relying on assumptions about proto-galactic magnetic fields.
However, it is important to note that the various conjectured origins of magnetic fields are not mutually exclusive, and future studies are required to distinguish their respective contributions to the production of intracluster magnetic fields.

\vspace{2ex}
In the final stage of this work, we became aware of a similar paper presenting results from an independent PIC simulation study of Weibel-seeded fluctuation dynamo in a pair plasma~\cite{Sironi2023}. In areas of overlap between our results and theirs, we find agreement.

\vspace{2ex}
\paragraph{Acknowledgements.}
The authors gratefully acknowledge A.~A.~Schekochihin and S.~C.~Cowley for insightful discussions. 
MZ thanks L.~Arzamasskiy for assistance with the collisionality measurement, J.~Mahlmann for help with data visualization, and P.~Kempski for discussions on magnetic field geometry. 
Support for MZ and MWK was provided by NSF CAREER award No.~1944972. Addition support for MZ was provided by a President's Fellowship from Princeton University. 
Support for DAU was provided by NASA grants 80NSSC20K0545 and 80NSSC22K0828, and NSF grant AST-1806084.
Support for VZ was provided by the Flatiron Institute. Research at the Flatiron Institute is supported by the Simons Foundation.
Support for NFL was provided by NSF CAREER Award 1654168.
High-performance computing resources supporting this work were provided by Texas Advanced Computing Center (TACC) at The University of Texas at Austin under Stampede2 allocation TG-PHY160032 and Frontera allocation AST20010. This work used the Extreme Science and Engineering Discovery Environment (XSEDE), which was supported by NSF grant number ACI-1548562.
\bibliography{ref}

\clearpage
\appendix

\setcounter{equation}{0}
\setcounter{figure}{0}
\setcounter{table}{0}
\newcounter{SIfig}

\renewcommand{\theequation}{S\arabic{equation}}
\renewcommand{\thefigure}{S\arabic{figure}}
\renewcommand{\thetable}{S\arabic{table}}
\renewcommand{\theSIfig}{S\arabic{SIfig}}

\section*{Supplementary Material}

\section{Details of Numerics.}

\subsection{External forcing}
\label{sec:forcing}
In our simulations, an external mechanical body force, $\bb{F}_{\rm ext}(\boldsymbol{x},t) = F_{{\rm ext},x}(\boldsymbol{x},t)\bb{\hat{x}}+F_{{\rm ext},y}(\boldsymbol{x},t)\bb{\hat{y}}+F_{{\rm ext},z}(\boldsymbol{x},t)\bb{\hat{z}}$,
is applied to particles to drive large-scale bulk flows.
The forcing is composed of a superposition of sinusoidal modes in space, having six solenoidal (shearing) modes at the box scale, with wavevectors $\bb{k}L/2\pi \in \{(0,1,0),(0,0,1)\}$ for $F_{{\rm ext},x}$, $\bb{k}L/2\pi \in \{(1,0,0),(0,0,1)\}$ for $F_{{\rm ext},y}$, and $\bb{k}L/2\pi \in \{(0,1,0),(1,0,0)\}$ for $F_{{\rm ext},z}$.
The amplitude for each mode of the force is chosen to be $T_{e0}/(\sqrt{6}L)$. Although the force is correlated at the large scale, it has a random phase at each value of $\bb{k}$ that is evolved independently following the Langevin equation in \citet{tenbarge2014oscillating}. To obtain a low Mach number flow, we choose a very low driving frequency $\omega_0$ and decorrelation rate $\gamma_0$ relative to the thermal timescale; specifically, $\omega_0=0.03 v_{\rm th}/(L/2\pi)$ and $\gamma_0=0.83\omega_0$.

\subsection{Inverse Compton (IC) radiation}
\label{sec:radiation}
In order to achieve a steady temperature, external IC radiative cooling is included in the simulations.
The emission process of IC radiation (in the optically thin limit) exerts a radiation backreaction force 
\begin{equation}
    \bb{F}_{\rm IC}= -\frac{4}{3} \sigma_{\rm T} U_{\rm ph} \gamma^2 \bb{v}/c
\end{equation}
to electrons and positions~\cite{Landau1975}. Here $\sigma_{\rm T} =(8 \pi/3)(e^2/m_e c^2)^2$ is the Thomson cross-section, $U_{\rm ph}$ is the energy density of the ambient photon field (with the photon density assumed to be isotropic), and $\gamma = (1-v^2/c^2)^{-1/2}$ is the particle Lorentz factor. 
In contrast to synchrotron cooling, which drives pressure anisotropy of the plasma by reducing the field-perpendicular component of the plasma pressure \citep{zhdankin_etal_2023}, the IC cooling is isotropic and mainly radiates at the high-energy tails of the plasma distribution.
Therefore, the IC cooling is not expected to affect the dynamics of the mirror and firehose instabilities, or the properties of the plasma dynamo. 

\subsection{Measurements of effective collisionality }
\label{sec:nu_eff_measurement}
The effective collisionality presented in the numerical results (in Figure 7) is quantified by the pitch-angle scattering rate and measured by studying the time evolution of magnetic moments $\mu$ of $10^4$ tracked particles.
We first divide the entire evolution into 20 time intervals.
In each time interval, we look at the time evolution of $\mu$ for each tracked particle, compute the collision time $\tau_{\rm coll}$ required for $\mu$ to change by a factor of $e$, and record the particle's gyro-radius $\rho_e$ averaged over all the time steps within the time interval $\tau_{\rm coll}$.
We then divide the ensemble of $\tau_{\rm coll}$ into three groups based on the associated $\rho_e$ ($\rho_e \geq L/2$, $L/30<\rho_e<L/2$, and $\rho_e \leq L/30$), and compute the histogram of $\tau_{\rm coll}$ for each group.

\begin{figure}[h]
    \centering
    \includegraphics[width=0.45\textwidth]{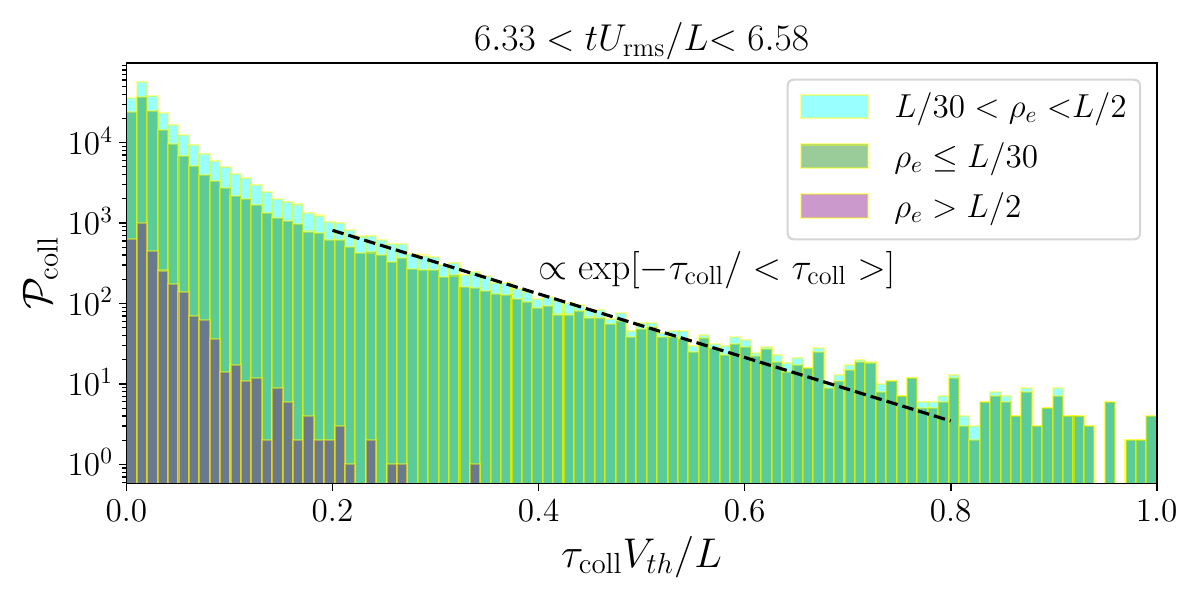}
    \caption{An example histogram of collision time $\tau_{\rm coll}$ of tracked particles, grouped according to their $\rho_e/L$, during the time interval $6.33<tU_{\rm rms}/L<6.58$. }
    \label{fig:tau_coll}
\end{figure}

An example histogram for the time interval $6.33<tU_{\rm rms}/L<6.58$ is shown in Figure~\ref{fig:tau_coll}. 
The $\tau_{\rm coll}$ of particles from the unmagnetized group $\rho_e \geq L/2$ cannot be used to calculate the scattering rate, because $\mu$ is not an adiabatic invariant for unmagnetized particles.
The histogram of the group $L/30<\rho_e<L/2$ is similar to that of the well-magnetized group $\rho_e \leq L/30$, suggesting that the particles are sufficiently magnetized to study the statistics of $\tau_{\rm coll}$.
The characteristic collision time $\braket{\tau_{\rm coll}}$ is obtained by fitting the histogram with an exponential function, $\exp{[-\tau_{\rm coll} /\langle \tau_{\rm coll} \rangle]}$, while the range $\tau_{\rm coll} V_{\rm th}/L < 0.2$ is not taken into account for the fitting to exclude the change of $\mu$ due to Bohm-like diffusion (i.e., particles sampling multiple field reversals during their gyromotion). 
The effective collisionality for each time interval is then evaluated as $\nu_{\rm eff}\equiv 1/\braket{\tau_{\rm coll}}$.

\section{Parameter scan of $L/d_e$}

\begin{figure}[h!]
    \centering
    \includegraphics[width=0.38\textwidth]{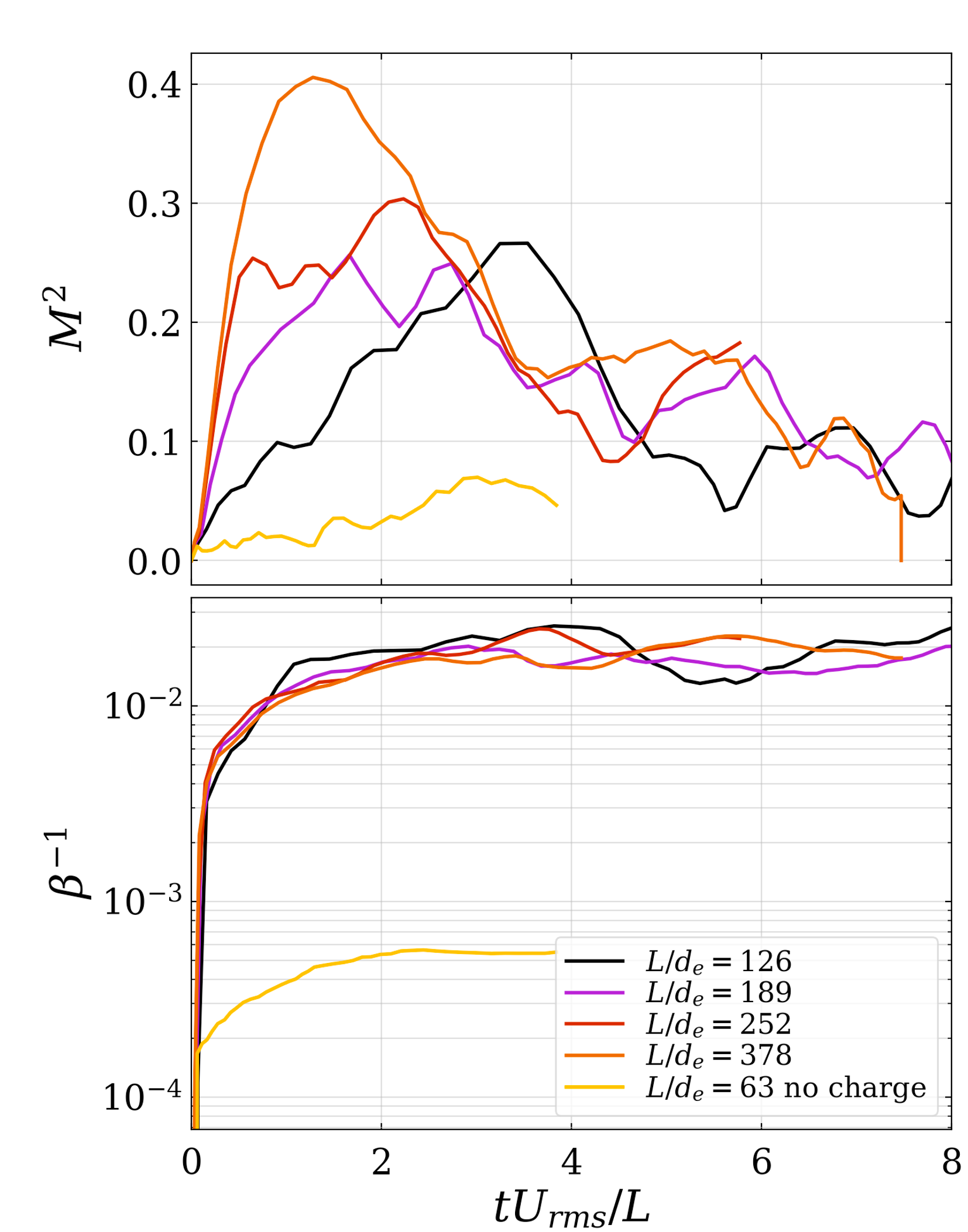}
    \caption{Time evolution of $M^2$ (top panel) and $\beta^{-1}$ (bottom panel) for runs with varying $L/d_e \in \{63,126,189,252,378\}$ and steady-state Mach number $M \approx 0.3$. The run with $L/d_e=63$ uses uncharged particles.}
    \label{fig:scanLde_S1}
\end{figure}

\begin{figure}[h!]
    \centering
    \includegraphics[width=0.38\textwidth]{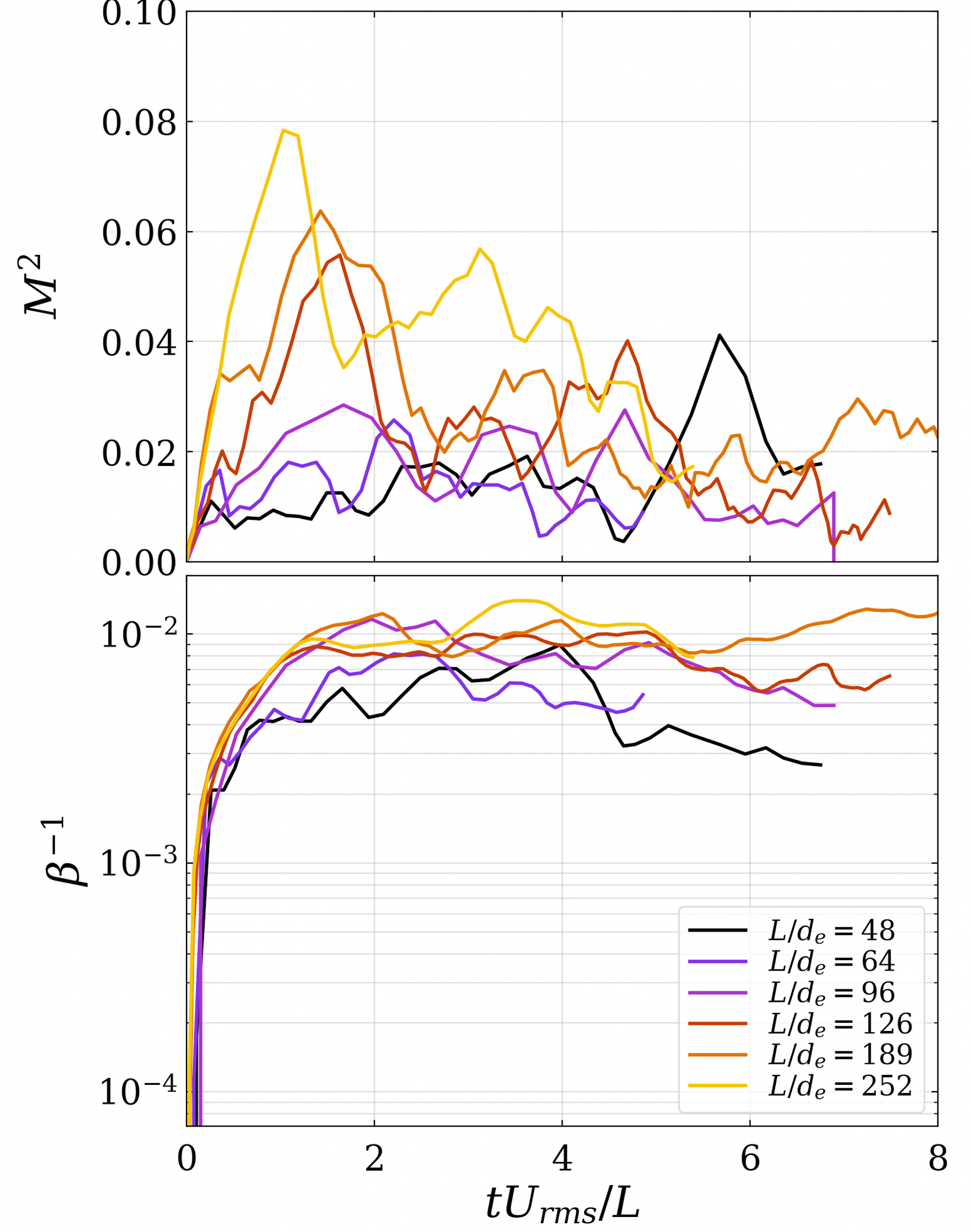}
    \caption{Time evolution of $M^2$ (top panel) and $\beta^{-1}$ (bottom panel) for runs with varying $L/d_e \in \{48,64,96,126,189,252\}$ and steady-state Mach number $M \approx 0.1$. }
    \label{fig:scanLde_Sp5}
\end{figure}

As a supplement to the main text, we present two group of runs with varying scale separation. The first group has $M \simeq 0.3$ at steady state and $L/d_e \in \{126,189,252,378\}$, as well as one run with uncharged particles (of which $L/d_e$ is an irrelevant parameter because $d_e$  has no physical meaning). 
The second group has $M \simeq 0.1$ and $L/d_e \in \{48,64,96,126,189,252\}$. 
The time evolution of $M^2$ and $\beta^{-1}$ for the first (second) group is shown in Figure~\ref{fig:scanLde_S1} (Figure~\ref{fig:scanLde_Sp5}). 

The run with uncharged particles provides a benchmark for the effects of magnetic fields.
As the system is continuously driven by the external force, the $\beta^{-1}$ stays at the level of numerical noise.
The $M^2$ increases slowly and is of a much smaller value than other runs, consistent with the argument that an unmagnetized plasma (or the neutral gas here) is subject to efficient phase mixing, and thus is effectively viscous. 

The time evolution of $M^2$ differs for systems with varying $L/d_e$. Within each group, although the values of $M^2$ at the steady state are similar, runs with larger $L/d_e$ have a faster acceleration of the flow in the beginning and shoot to a higher value of $M^2$ before decreasing to the steady value. 
This is consistent with the argument made in the main text that during the initial Weibel phase, the effective collisionality is determined by the particle scattering at the ends of the Weibel filaments.
With larger $L/d_e$, the Weibel filaments have smaller length scales in each dimension, which lead to a shorter mean free path of the particles, i.e., larger effective collisionality or smaller viscosity for the system, and thus a faster acceleration and higher peak $M$ for the flows. However, it seems unlikely that this trend will continue to values of $M > 1$, when the flow becomes supersonic.

Systems with varying $L/d_e$ (for $L/d_e \gtrsim 100$) show similar evolution of $\beta^{-1}$ (Figure~\ref{fig:scanLde_S1}, bottom panel).
The levels of $\beta^{-1}$ given by the Weibel seed fields have a weak dependence on $L/d_e$, as expected given Eq.~(1).
The subsequent plasma dynamo amplifies the magnetic fields with a growth rate that is similar for all these runs.
This is consistent with the fact that even for the run with the largest $L/d_e$ (=378), which is analyzed in the main text, the parallel rate of strain is mainly given by flows at the forcing scale.
The dynamo growth rate (for runs with $L/d_e\le 378$) is tied to the flow-crossing rate (at the forcing scale), independent of $L/d_e$.
For the run with $L/d_e=48$ (Figure~\ref{fig:scanLde_Sp5}, bottom panel), it is unclear whether a dynamo phase exists after the Weibel stage. This could be due to the significant electron Landau damping of the magnetic fields under the very limited scale separation.  
This suppression of dynamo at small system sizes is consistent with the previous study demonstrating the role of electron Landau damping in inhibiting dynamo~\cite{pusztai2020dynamo}.

\end{document}